\documentclass[
aps,%
prd,
10pt,%
notitlepage,%
oneside,
nobibnotes,%
nofootinbib,%
superscriptaddress,%
noshowpacs,%
centertags]%
{revtex4-2}
\usepackage[cp1251]{inputenc}
\usepackage[T2A]{fontenc}
\usepackage{epsf,amsmath,amsfonts,amssymb,amsbsy}
\usepackage[mathscr]{eucal}
\usepackage{hyperref}
\usepackage{graphicx}

\begin{document}
	
\title{Causal Space-Time Structure\\ and non-Hausdorff Extension of Schwarzschild Black Hole Interior}
	
\begin{abstract}
Cones of future and past are rigorously constructed under the horizon of black hole by completions of geodesics continued from causally connected space-time regions outside the black hole interior. Treating the Schwarzschild black hole as a zero charge limit of Reissner--Nordstr\o m black hole reforms the Carter--Penrose diagram into the infinite band that excludes the closed proper-time cycles under the  geodesic motion. The marginal trajectories between the cones of future and past compose paths common with a compact non-Hausdorff spherical extension of black hole interior forming the thermal bath for particles fallen into the black hole.
\end{abstract}

	\author{\firstname{Asya}~\surname{Aynbund}}
	\email{aynbund.asya@phystech.edu}
	\affiliation{Landau Phystech-School, Moscow Institute of Physics and Technology,
		Russia, 141701, Moscow Region, Dolgoprudny, Institutsky 9}
	
	\author{V.V.Kiselev}
	\email{kiselev.vv@phystech.edu; Valery.Kiselev@ihep.ru}
	
	\affiliation{Landau Phystech-School, Moscow Institute of Physics and Technology,
		Russia, 141701, Moscow Region, Dolgoprudny, Institutsky 9} 
	\affiliation{Institute for High Energy Physics named by A.A.\,Logunov of National
Research Centre <<Kurchatov Institute>>, 
		Russia, 142281, Moscow Region, Protvino, Nauki 1}
	
	\maketitle
\section{Introduction}	
Investigating various black hole solutions in general relativity is a tool to touch extraordinary conditions in interactions between the gravity and matter. These studies are important especially if one considers a causal structure of motion, which can open unexpected quantum effects such as the Hawking radiation and the information paradox, for instance. Recently in \cite{tHooft:2021lyt,tHooft:2024auh} G.\,'t Hooft has freshened up such studies by some novel ideas particularly including a  statement of doubling the Hawking temperature of Schwarzschild black hole. In \cite{Morozov:2023jsp} Alexey Yu.\,Morozov also has published original speculations on the causal structure of information paradox under forming  portals to new Universes in the black hole interior. Some fresh ideas about the singularity inside the Schwarzschild black hole have been published by  D.-f.\,Zeng in \cite{Zeng:2025lch} after we independently completed the research given here: motivations, results and conclusions in both papers are complementary each to other (see also \cite{Zeng:2023ueq}).

The present paper is devoted to precise the causal space-time structure under the radial motion of massive particle in the black hole interior in the Hamilton--Jacobi equation framework for time-like trajectories in the Schwarzschild black hole space-time. New aspects appear by treating a reflection of particle motion at turn points inside the black hole interior under conserving an integral of motion. This approach results in a principal modification concerning specification of correct cones for causal future and past. At the same time an admissible change in the integral of motion opens a non-Hausdorff extension of black hole interior. 

To explicitly show the problem setting let us consider the Hamilton--Jacobi equation for a free radial motion of massive particle with action $S$ and mass $m$
\begin{equation}
\label{in1}
	g^{\mu\nu}\,\big(\partial_\mu S\big)\,\big(\partial_\nu S\big)=m^2
\end{equation}
in the Schwarzschild metric
\begin{equation}
\label{in2}
	\mathrm ds^2=g_{\mu\nu}\,\mathrm d x^\mu\mathrm d x^\nu =
	\left(1-\frac{r_g}{r}\right)\mathrm d t^2-\frac{\mathrm d r^2}{1-\frac{r_g}{r}}
	-r^2\,\big(\mathrm d\theta^2+\sin^2\theta\,\mathrm d \phi^2\big),\qquad r_g=2MG,\quad
	g_{tt}=1-\frac{r_g}{r},
\end{equation}
written down in spherical coordinates of distant observer.  Here $r_g$ is the Schwarzschild black hole radius,  $M$ is the black hole mass, $G$ is the Newton constant. The radial motion sets $\mathrm d \theta=\mathrm d \phi =0$, while the temporal Killing vector $\partial_t$ involves the energy as the integral of motion
\begin{equation}
\label{in3}
	\partial_t S=-E=\mbox{const.}
\end{equation}
Then the Hamilton--Jacobi equation for the action at a radial time-like trajectory under the fixed integral of motion 
\begin{equation}
\label{in4}
	S=-E\,t+\mathscr S(r)
\end{equation}
in accordance with (\ref{in1}) takes the form 
\begin{equation}
\label{in5}
	\frac{E^2}{1-\frac{r_g}{r}}-\left(1-\frac{r_g}{r}\right)\big(\partial_r\mathscr S\big)^2=m^2\quad\Rightarrow\quad
	\frac{E^2}{m^2}=g_{tt}+\frac{1}{m^2}\left(g_{tt}\,\partial_r\mathscr S\right)^2.
\end{equation}
Introducing so called <<tortoise coordinate>> $r_\star$ by
\begin{equation}
\label{in6}
	\mathrm d r_\star=\frac{\mathrm d r}{g_{tt}}\quad \Rightarrow\quad
	r_\star =r+r_g \ln\left(\frac{r}{r_g}-1\right),\qquad g_{tt}\,\partial_r\mathscr S=\partial_{r_\star}\mathscr S
\end{equation}
and <<turn radius>> $r_c$ as the integral of motion
\begin{equation}
\label{in7}
	\frac{E^2}{m^2}=1-\frac{r_g}{r_c} \quad \Rightarrow\quad r_c=\frac{r_g}{1-{E^2}/{m^2}}
\end{equation}
we rewrite the previously transformed Hamilton--Jacobi equation  (\ref{in5}) for the radial motion into the ordinary equation of energy conservation in mass units
\begin{equation}
\label{in8}
	\frac{E^2}{m^2}=g_{tt}+\frac{1}{m^2}\big(\partial_{r_\star}\mathscr S\big)^2 \quad\Leftrightarrow\quad
	1-\frac{r_g}{r_c} =1-\frac{r_g}{r}+\frac{1}{m^2}\big(\partial_{r_\star}\mathscr S\big)^2\quad\Rightarrow\quad
	r_g\,\frac{r_c-r}{r_c r}=\frac{1}{m^2}\big(\partial_{r_\star}\mathscr S\big)^2,
\end{equation}
wherein the sum of <<potential>> $g_{tt}$  and kinetic term $\big(\partial_{r_\star}\mathscr S\big)^2/m^2$ remains constant. 
Note, that the differential $\mathrm d r_\star$ is real at any real $r>0$. So, the derivative of action  $\partial_{r_\star}\mathscr S$ is also real, hence, the right hand sides of equations in (\ref{in8}) are well defined at $r_c\geqslant r>0$, too. According to (\ref{in6}) differentials $\mathrm d r_\star$ and $\mathrm d r$ are real at all of time-like geodesics with positive intervals $\mathrm ds^2>0$. 

\begin{figure}[t]
\begin{center}
\includegraphics[width=8.5cm]{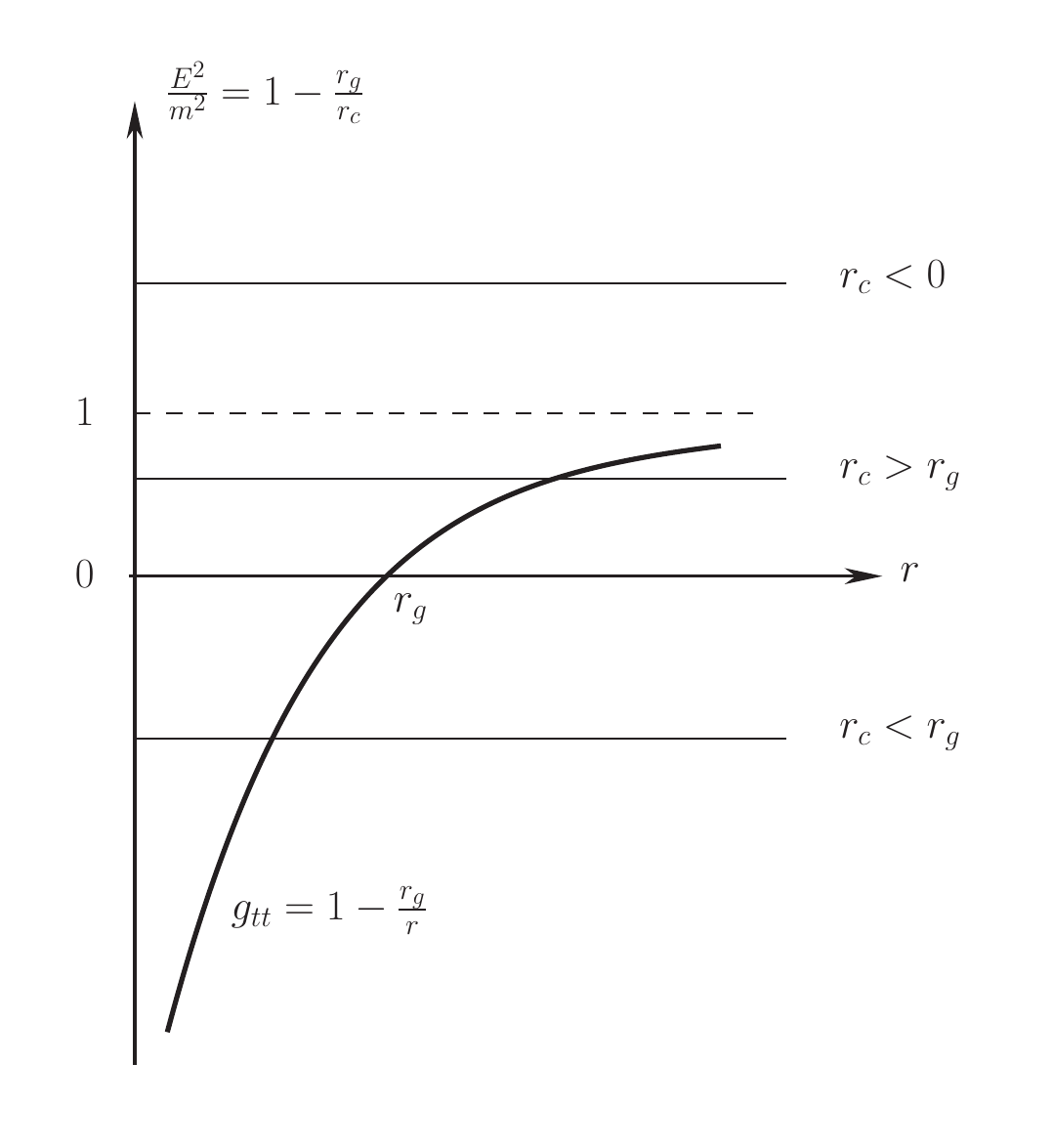}
\caption{Different levels of motion invariant expressed in $r_c$ parameter and the sum of <<potential>> and <<kinetic>> terms in (\ref{in8}).}
\label{H-M-fig}
\end{center}
\end{figure}

At $E>0$ one gets $r_c>r_g$ and $r_c$ determines the maximal radial coordinate, i.e. the turn point of radial motion, indeed. 
Such a situation is shown in FIG. \ref{H-M-fig}, when the turn point of massive particle is posed in the black hole exterior. Formally, at $r_c<0$ the massive particle reaches the distant infinity $r\to +\infty$ under a velocity $v_\infty<c$, since the action
\begin{equation}
\label{in9}
	S=-m\int\mathrm d s=-Et+\int\limits^r\mathrm d \mathscr S(r)
	=-Et+\int\limits^{r_\star(r)}\mathrm d r_\star\,\frac{\partial \mathscr S(r)}{\partial r_\star}
	=-Et+m\int\limits^{r}\frac{\mathrm d r}{g_{tt}}
	\sqrt{\frac{E^2}{m^2}-g_{tt}}
\end{equation}
determines the time-like interval $\mathrm d s^2 >0$. Let us prove it. So, according to the Hamilton--Jacobi framework the derivative of action with respect to the integral of motion holds constant. Therefore,
\begin{equation}
\label{in11}
	\frac{\partial S}{\partial E}=\mbox{const.}\quad \Rightarrow\quad
	Et=\frac{E^2}{m^2}\int\limits^r\frac{\mathrm d r}{g_{tt}}
	\frac{m}{\sqrt{\frac{E^2}{m^2}-g_{tt}}}\quad \Rightarrow\quad
	\mathrm d S=-E\mathrm d t+\mathrm d \mathscr S(r)=-m\,
	\frac{\mathrm d r}{\sqrt{\frac{E^2}{m^2}-g_{tt}}}.
\end{equation}
Thus, at $r > 0$ and $1/r > 1/r_c$, hence, at $r < r_c$ for $r_c > 0$ and at any $r$ for $r_c < 0$,
\begin{equation}
\label{in10}
	\mathrm d s^2 =\mathrm d r^2\,\frac{r_c}{r_g}\,\frac{r}{r_c-r}>0.
\end{equation}
It is remarkable that the interval remains time-like even at $r_c\leqslant r_g$, when the trajectory is inside the black hole interior. In FIG. \ref{H-M-fig} this class of radial trajectories corresponds to the region below the abscissa axis. 

At radial geodesics the proper time $\tau$ of massive particle satisfies
$$
	\mathrm d \tau^2\equiv\mathrm d s^2= g_{tt}\mathrm d t^2-\frac{1}{g_{tt}}\,\mathrm d r^2>0,
$$
which together with (\ref{in10}) results in 
\begin{equation}
\label{in11-a}
	\mathrm d s^2=\frac{r_c}{r_c-r_g}\,g_{tt}^2\,\mathrm d t^2> 0\qquad\mbox{at } r_c>r_g.
\end{equation}
Therefore, at time-like radial trajectories with $r_c> r_g$ or $r_c<0$ the differential $\mathrm d t$ is well defined and it is real even at $r<r_g$ behind the horizon, while for the integral of motion inside $0<r_c<r_g$ to be lying behind the horizon there is no any notion of time differential for time-like geodesics though the proper time is correctly defined in (\ref{in10}).

In Section \ref{sec-nonH} we show that radial time-like trajectories possessing the motion integral within $0<r_c\leqslant r_g$ are naturally associated with the non-Hausdorfian extension of black hole interior, and this extension is continuously connected to two-dimensional Kruskal plane along common tangent lines. 

At the trajectory the interval equals 
\begin{equation}
\label{in12}
	\mathrm d s^2 =g_{tt}\big(\mathrm d t^2-\mathrm d r_\star^2\big)=
	g_{tt}\mathrm d r_\star^2\left(\left(\frac{\mathrm d t~}{\mathrm d r_\star}\right)^2-1\right)=
	\mathrm d r_\star\mathrm d r \left(\left(\frac{\mathrm d t~}{\mathrm d r_\star}\right)^2-1\right),
\end{equation}
where the quantity 
\begin{equation}
\label{in13}
	\left(\frac{\mathrm d t~}{\mathrm d r_\star}\right)^2=
	\frac{r_c-r_g}{r_g}\,\frac{r}{r_c-r}>0 \qquad\mbox{at }r_c>r_g
\end{equation}
and its inverse value are well determined from (\ref{in10}) as the function of radial coordinate $r$ in the whole interval of $r$ in the trajectory at $r_c\geqslant r_g$, while the singular point is posed at $r=0$. The <<tortoise velocity>> $\mathrm d r_\star/\mathrm d t$ changes its sign at the turn point given by $r_\mathrm{turn}=r_c>0$ wherein this velocity vanishes $\mathrm d r_\star/\mathrm d t=0$. 

Let us define the reflection point $r_\mathrm{refl}$ for the radial geodesics with $r_c\geqslant r_g$  by the same condition: the reflection conserves the integral of motion and changes the sign of radial velocity $\mathrm d r_\star/\mathrm d t$. We apply this definition in order to specify causal cones of future and past in Section \ref{sec-reflect}. We show that the reflection method is universal: it is legal regardless whether the trajectory is considered in the black hole exterior or interior. This universality allows us to precise a construction of causal cones, that  supports a treatment of $r=0$ as the turn point reflecting the trajectory into a complimentary continuation of space-time as an infinite band excluding any return to the former past. 

The results are summarised and discussed in Conclusion.

\section{Reflection points and setting the cones of future and past\label{sec-reflect}}
For the case of $r_c>r_g$ one can use the transform of space-time coordinates from the time $t$ and distance $r$ of distant observer at $r>r_g$ to the retarded and advanced time $u$ and $v$ set by 
\begin{equation}
\label{2in-1}
	\left\{\begin{array}{l}
	u = t-r_\star,\\ v=t+r_\star,
	\end{array}
	\right.\qquad r_\star =r+r_g\ln\left(\frac{r}{r_g}-1\right),
\end{equation}
from which the Kruskal variables $\bar u$ and $\bar v$ are build
\begin{equation}
\label{eq:kruskal}
	\left\{\begin{array}{l}\displaystyle 
	\bar u = -2 r_g\,\exp\left(-\frac{u}{2r_g}\right),\\ [5mm] \displaystyle 
	\bar v = +2 r_g\,\exp\left(\frac{v}{2r_g}\right),
	\end{array}
	\right.\qquad \Leftrightarrow\qquad
	\left\{\begin{array}{ll}
	t &\hskip-2.mm=\displaystyle  r_g\,\ln\left(-\frac{\bar v}{\bar u}\right),\\[5mm]  
	r_\star &\hskip-2.mm=\displaystyle  r_g\,\ln\left(-\frac{\bar v\cdot \bar u}{4r_g^2}\right).
	\end{array}
	\right.
\end{equation}
At $r < r_g$ ($\bar u \bar v > 0$) the logarithms in \eqref{eq:kruskal} are defined up to the imaginary constant $\mathrm{i}\pi r_g$, which shifts $t$ and $r_\star$ but not their differentials, so we can omit this constant. The metrics on the $(\bar v,\bar u)$-plane takes the form
\begin{equation}
\label{ds-uv}
	\mathrm d s^2=\frac{r_g}{r}\,\mathrm e^{-r/r_g}
	\cdot\mathrm d \bar v\,\mathrm d \bar u.
\end{equation}
So, light geodesics are determined by straight lines $\bar v=\mbox{const.}$ or $\bar u=\mbox{const.}$
The horizon $r=r_g$ is described by two lines $\bar v =0$ or $\bar u=0$. 
The singularity $r=0$ is the following hyperbola:
$$
	\frac{\bar v\cdot \bar u}{4r_g^2}=1,
$$
while the region of $\bar u>0$ and $\bar v>0$ refers to the black hole and 
the region of $\bar u<0$ and $\bar v<0$ refers to the white hole of stationary Schwarzschild solution for the distant observer posed at $\bar v>0$ and $\bar u<0$. 

Differentials in (\ref{eq:kruskal}) give
\begin{equation}
\label{2in-3}
	\left\{\begin{array}{ll}\displaystyle 
	\frac{1}{r_g}\,\mathrm d t &\hskip-2.mm= \displaystyle 
	\frac{\mathrm d \bar v}{\bar v}-\frac{\mathrm d \bar u}{\bar u},\\[5mm] \displaystyle 
	\frac{1}{r_g}\,\mathrm d r_\star &\hskip-2.mm= \displaystyle 
	\frac{\mathrm d \bar v}{\bar v}+\frac{\mathrm d \bar u}{\bar u}.
	\end{array}
	\right.
	\qquad \Rightarrow\qquad
	\frac{\mathrm d t~}{\mathrm d r_\star} = 
	\frac{\bar u\mathrm d \bar v-\bar v\mathrm d \bar u}
	{\bar u\mathrm d \bar v+\bar v\mathrm d \bar u}=
	\frac{\bar u-\bar v\,\frac{\mathrm d \bar u}{\mathrm d \bar v}}
	{\bar u+\bar v\,\frac{\mathrm d \bar u}{\mathrm d \bar v}}
	\qquad \Rightarrow\qquad
	\frac{\mathrm d \bar u}{\mathrm d \bar v}=\frac{\bar u}{\bar v}\,
	\frac{1-\frac{\mathrm d t~}{\mathrm d r_\star}}{1+\frac{\mathrm d t~}{\mathrm d r_\star}}.
\end{equation}
We see that all differentials are real values. It is important to emphasize that quantity ${\mathrm d t}/{\mathrm d r_\star}$ and its inverse are well defined as the function of geodesics on the $(\bar v, \bar u)$-plane.

At the reflection point of some fixed Kruskal-plane point on the geodesics the velocity changes the sign or
$$
	\frac{\mathrm d t~}{\mathrm d r_\star}\;\mapsto\; -\frac{\mathrm d t~}{\mathrm d r_\star},
$$
hence, the tangent lines before the reflection (prior) and after the reflection (post) satisfy
\begin{equation}
\label{2in-3+}
	\frac{\mathrm d\ln \bar u}{\mathrm d\ln \bar v}\Bigg|_\mathrm{prior}=
	\frac{1-\frac{\mathrm d t~}{\mathrm d r_\star}}{1+\frac{\mathrm d t~}{\mathrm d r_\star}}\;
	\mapsto\; \frac{1+\frac{\mathrm d t~}{\mathrm d r_\star}}{1-\frac{\mathrm d t~}{\mathrm d r_\star}}
	=\frac{\mathrm d\ln \bar u}{\mathrm d\ln \bar v}\Bigg|_\mathrm{post}
	\qquad \Rightarrow\qquad
	\frac{\mathrm d\ln \bar u}{\mathrm d\ln \bar v}\Bigg|_\mathrm{prior}=
	\frac{\mathrm d\ln \bar v}{\mathrm d\ln \bar u}\Bigg|_\mathrm{post}.
\end{equation}
Geometrically, the reflection reverses the velocity relative to the static observer at the point, whose worldline is the curve $r = \mathrm{const}$. Denoting the direction to the position of reflection point from the center on the $(\bar v,\bar u)$-plane  as
$$
	\frac{\bar u}{\bar v}\equiv \tan \alpha
$$
and the direction of motion as 
$$
	\frac{\mathrm d \bar u}{\mathrm d \bar v}\Bigg|_\mathrm{prior}=\tan\alpha_\mathrm{prior}
	\qquad\mbox{and}\qquad 
	\frac{\mathrm d \bar u}{\mathrm d \bar v}\Bigg|_\mathrm{post}=\tan\alpha_\mathrm{post},
$$
we find from (\ref{2in-3+}) the following expression for the direction after the reflection:
\begin{equation}
\label{eq:reflection}
	\tan\alpha_\mathrm{post}=\frac{\tan^2\alpha}{\tan\alpha_\mathrm{prior}}.
\end{equation}

Note that $\alpha=\mbox{const.}$ represents the straight line of constant time in accordance with (\ref{eq:kruskal}), and if $t_2>t_1$ then for negative angles $\alpha_2>\alpha_1$.

Since a curve of constant radius $r=\mbox{const.}$ means a constant $r_\star$ the connection in (\ref{in6}) defines the tangent line to the curve as
$$
	\bar u\,\mathrm d \bar v_\|+\bar v\,\mathrm d \bar u_\|=0 \qquad \Rightarrow\qquad
	\frac{\mathrm d \bar u_\|}{\mathrm d \bar v_\|}\equiv \tan\alpha_\|=-\frac{\bar u}{\bar v}
	=-\tan \alpha,
$$
hence, $\alpha_\|=-\alpha$ and 
\begin{equation}
\label{eq:reflection-par}
	\tan\alpha_\mathrm{post}=\frac{\tan^2\alpha_\|}{\tan\alpha_\mathrm{prior}}.
\end{equation}

\subsection{Domain of distant observer}
\begin{figure}[h]
\begin{center}
\includegraphics[width=10.5cm]{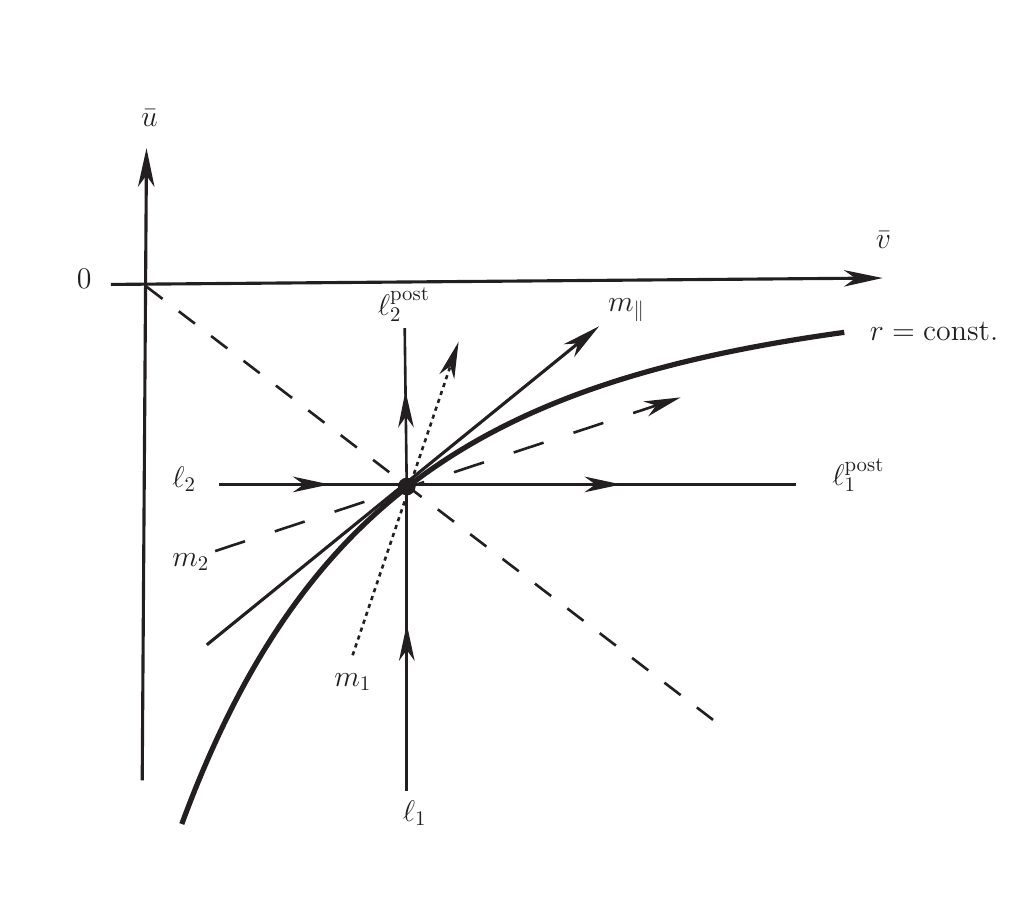}
\caption{The point of possible reflections for various radial geodesics in the distant observer quadrant of $(\bar v,\bar u)$-plane. Notations: tangent line $m_\|$ for $r=\mbox{const.}$, light lines marked by $\ell$, massive lines marked by $m$.}
\label{pic1}
\end{center}
\end{figure}

Consider the distant observer quadrant set by $\bar v>0$ and $\bar u<0$ as shown in FIG. \ref{pic1}. The tangent line for a point lying on the curve $r=\mbox{const.}\equiv r_0$ is denoted by symbol $m_\|$. If the curve corresponds to maximal radius $r_c=r_0>r_g$, then the point is the turn point, indeed, and the massive particle arrives to the point and leaves it toward the arrow in the $m_\|$ line, the velocity vanishes at the turn point and changes its sign continuously. So, $m_\|$ corresponds to the case when the reflection point $r_0$ coincides with the turn point $r_c$, while $\alpha_\mathrm{prior}=\alpha_\|$ and (\ref{eq:reflection-par}) yields $\alpha_\mathrm{post}=\alpha_\|$. 
Line $m_1$ passes through the point in two cases: the particle moves from the infinity or from $r_c>r_0$ since the line depends not only on the direction, but also on the absolute value of velocity. Since \eqref{eq:reflection-par} gives $\tan\alpha_{\rm prior}\cdot\tan\alpha_{\rm post} = \tan^2\alpha_\parallel$, the reflected line lies on the other side of $m_\parallel$: if $\tan\alpha_{\rm prior} > \tan\alpha_\parallel$, then $\tan\alpha_{\rm post} < \tan\alpha_\parallel$, and the particle is reflected towards the direction of $m_2$.  The same is valid for the massive particle moving along $m_2$, when $\tan\alpha_\mathrm{prior}<\tan\alpha_\|$, hence, (\ref{eq:reflection-par}) yields $\tan\alpha_\mathrm{post}>\tan\alpha_\|$, changing the resulting markers $1\leftrightarrow 2$. If $m_1$ tends to the light line $\ell_1$ one gets $\ell_2^\mathrm{post}$ for passing and $\ell_1^\mathrm{post}$ for reflecting. 

Thus, the cone of future is posed from $\ell_2^\mathrm{post}$ clockwise to $\ell_1^\mathrm{post}$, while the cone of past is from $\ell_1$ clockwise to $\ell_2$. Note, the particles falling into the black hole occupy the sector from $\ell_2^\mathrm{post}$ clockwise to $m_\|$ in the cone of future, while the particles falling out of the hole populate the sector from $m_\|$ clockwise to $\ell_1^\mathrm{post}$  in the cone of future. In both cases $(\mathrm d r_\star/\mathrm d t)^2<1$ due to (\ref{in12}). The velocity is negative for falling in, $\mathrm d r_\star/\mathrm d t<0$, and positive for falling out, $\mathrm d r_\star/\mathrm d t>0$. 

Let us prove  that these signs do not change, when crossing the horizon. According to (\ref{2in-3})
$$
	\frac{\mathrm d t~}{\mathrm d r_\star} = 
	\frac{\bar u\mathrm d \bar v-\bar v\mathrm d \bar u}
	{\bar u\mathrm d \bar v+\bar v\mathrm d \bar u},
$$
hence, falling in corresponds to crossing the horizon at  $\bar u =0$ yielding
\begin{equation}
\label{fall-in}
	\frac{\mathrm d t~}{\mathrm d r_\star} =-1,
\end{equation}
while falling out corresponds to crossing the horizon at  $\bar v =0$ yielding
\begin{equation}
\label{fall-out}
	\frac{\mathrm d t~}{\mathrm d r_\star} =+1.
\end{equation}
Between reflections, $\mathrm{d}r_\star/\mathrm{d}t$ is a smooth function of $r$ given by \eqref{in13}, and it equals $\pm 1$ at the horizon, so its sign cannot change there. Moreover, according to (\ref{in12}) the interval is time-like, that means $(\mathrm d r_\star/\mathrm d t)^2<1$ at $r>r_g$, when $g_{tt}>0$ and $(\mathrm d r_\star/\mathrm d t)^2>1$ at $r<r_g$, when $g_{tt}<0$. 

\begin{figure}[h]
\begin{center}
\includegraphics[width=5cm]{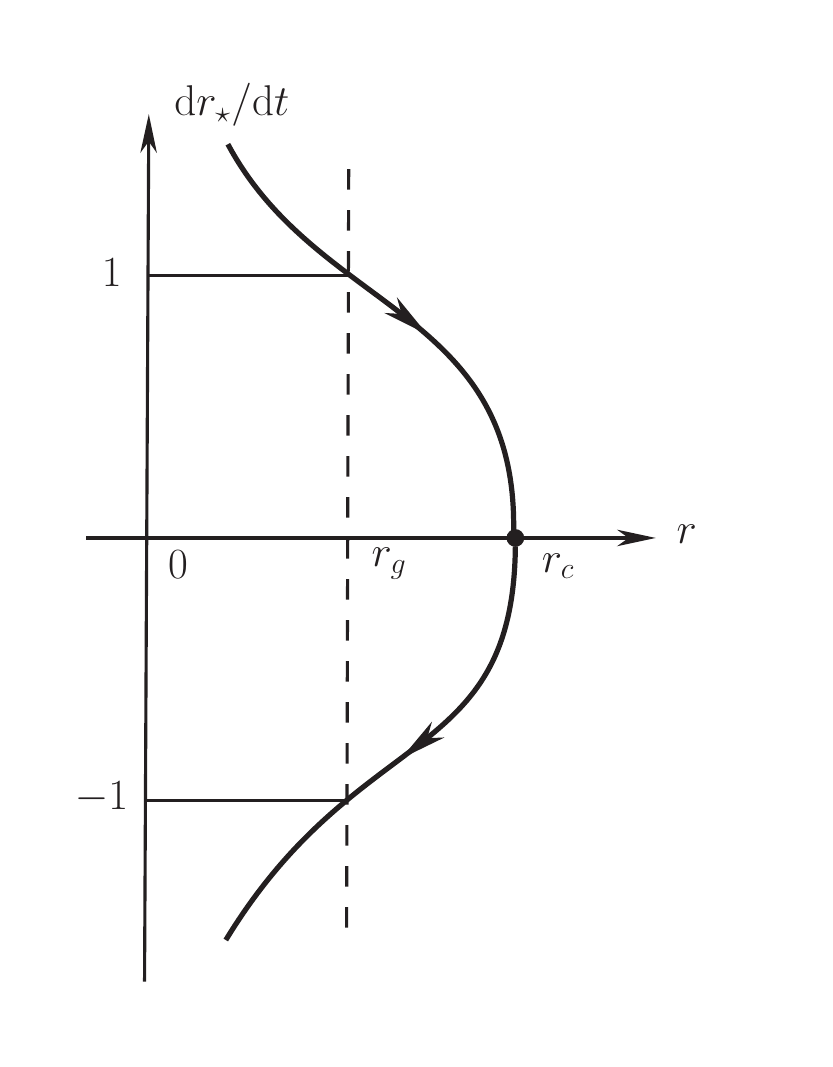}
\caption{The velocity $\mathrm{d}r_\star/\mathrm{d}t$ along the trajectory with $r_c > r_g$: it equals $\pm 1$ at the horizon $r = r_g$ and vanishes at the turn point $r = r_c$. }
\label{pic3}
\end{center}
\end{figure}

Therefore, we can draw the complete trajectory for $r_c>r_g$ as shown in FIG. \ref{pic3}. Similar picture appears for the case, when the trajectory falls into the black hole from infinity or falls out of the black hole to infinity.

This consideration with the usage of reflections of geodesics confirms the standard representation for cones of future and past in the distant observer domain. 

\subsection{Black hole interior}
\begin{figure}[t]
\begin{center}
\includegraphics[width=11cm]{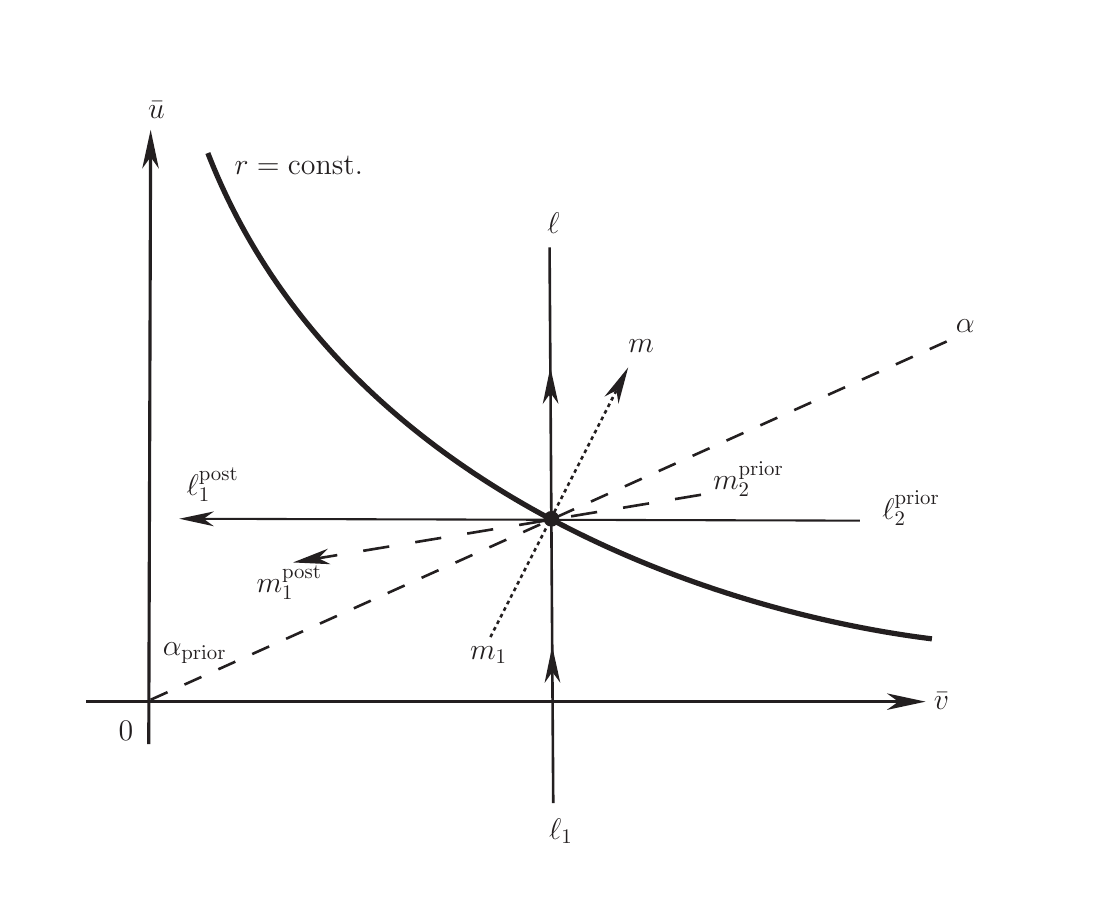}
\caption{The point of possible reflections for various radial geodesics in the black hole quadrant of $(\bar v,\bar u)$-plane. Notations: light lines marked by $\ell$, massive lines marked by $m$.}
\label{pic2}
\end{center}
\end{figure}

The domain inside the black hole is determined by quadrant set by $\bar v>0$ and $\bar u>0$ pictured in FIG. \ref{pic2}. In this domain the curve with constant radius $r_0=\mbox{const.}$ is posed at $r_0<r_g$ while the particle has $r_c>r_g$. So, there is no sense to consider the tangent line to the curve $r_0=\mbox{const.}$, since it cannot be associated with any turning of particle trajectory, and only reflections are meaningful.

Let us consider a marginal geodesic in the limit  $r_c=r_g+0$, hence, $E\to +0$ in accordance with (\ref{in7}). Relation (\ref{in13}) yields ${\mathrm d t}/{\mathrm d r_\star}\to 0$, and (\ref{2in-3}) determines the direction of motion
\begin{equation}
\label{2in-6}
	\frac{\mathrm d \bar u}{\mathrm d \bar v}=\frac{\bar u}{\bar v}.
\end{equation}
Therefore, the particle moves along the direction from the center to the point in the $(\bar v, \bar u)$-plane. Formally, at this geodesic $\tan\alpha_\mathrm{prior}=\tan\alpha$, hence, according to (\ref{eq:reflection}) the reflection gives the same direction $\tan\alpha_\mathrm{post}=\tan\alpha=\tan\alpha_\mathrm{prior}$. It means also that all of geodesics possessing $r_c>r_g$ and  falling in the black hole are posed between this marginal geodesic contra-clockwise to the light line  $\ell_1$ shown in FIG. \ref{pic2}. The light line $\ell_1$ without any reflection is continued to $\ell$ after the point of $(\bar v,\bar u)$-plane under consideration. Direction $m_1$ of massive particle falling in the black hole is also continued to $m$ if there is no reflection. Therefore, the evident part of future cone is covered by the sector from light line $\ell$ clockwise to the line marked by $\alpha$. 

The reflection of line $m_1$ with $\tan\alpha_{m_1}^\mathrm{prior}>\tan\alpha$ results in 
$$
	\tan\alpha_{m_1}^\mathrm{post}=
	\frac{\tan^2\alpha}{\tan\alpha_{m_1}^\mathrm{prior}}<\tan\alpha.
$$
We denote this line as $m_1^\mathrm{post}$. The question is whether the direction arrow is correct. 

According to (\ref{in11-a}) the proper time interval $\mathrm d \tau$ is given by 
\begin{equation}
\label{2in-7}
	\mathrm d \tau =g_{tt}\,\mathrm d t\cdot\sqrt{\frac{r_c}{r_c-r_g}},
\end{equation}
and it remains positive during any evolution of massive particle: $\mathrm d \tau>0$. Then $\mathrm d t>0$ at $r>r_g$ and  $\mathrm d t<0$ at $r<r_g$. Under definition of Kruskal variables in (\ref{2in-1}) and (\ref{eq:kruskal}) we find differentials
\begin{equation}
\label{eq:differentials}
	\left\{\begin{array}{ll}\displaystyle 
	\mathrm d \bar v &\hskip-2.mm= \displaystyle +\bar v\,\mathrm d t 
	\left(1+\frac{\mathrm d r_\star}{\mathrm d t~}\right)\cdot \frac{1}{2r_g},\\[5mm] \displaystyle 
	\mathrm d \bar u &\hskip-2.mm= \displaystyle -\bar u\,\mathrm d t 
	\left(1-\frac{\mathrm d r_\star}{\mathrm d t~}\right)\cdot \frac{1}{2r_g}.
	\end{array}
	\right.
\end{equation}
Recall that falling in corresponds to $\mathrm{d}r_\star/\mathrm{d}t < -1$; together with $\mathrm{d}t < 0$ behind the horizon, \eqref{eq:differentials} gives $\mathrm{d}\bar v > 0$ and $\mathrm{d}\bar u > 0$. After the reflection the position in the $(\bar v, \bar u)$-plane is not changed as well as the sign of proper time evolution does not change, too, while  $\mathrm d r_\star/\mathrm d t$ flips its sign at the same modulus, so that, $\mathrm d r_\star/\mathrm d t>1$. Since $\mathrm{d}t$ is unchanged and both factors $1 \pm \mathrm{d}r_\star/\mathrm{d}t$ in \eqref{eq:differentials} change sign, both differentials $\mathrm{d}\bar v$ and $\mathrm{d}\bar u$ change sign and become negative. Then line $m_1$ is reflected to $m_1^\mathrm{post}$, indeed. In the limit of tending the geodesic to the light line $m_1\to \ell_1$ we find the reflection of $\ell_1$ to light line $\ell_1^\mathrm{post}$ falling out from the black hole to the quadrant with $\bar v<0$ and $\bar u>0$ for another distant observer. 

Thus, the complete cone of future inside the black hole is composed of two sectors over the marginal geodesic of $\alpha_\mathrm{prior}=\alpha$, so it occupies the sector from $\ell$ clockwise to $\alpha$ and the sector from $\alpha_\mathrm{prior}$ clockwise to $\ell_1^\mathrm{post}$. 

The light line $\ell_1^\mathrm{post}$ is continuation of $\ell_2^\mathrm{prior}$ line that describes the light falling out from the black hole quadrant without any reflection, while $m_1^\mathrm{post}$ is continuation of $m_2^\mathrm{prior}$ that describes the massive particle geodesics falling out from the black hole quadrant without any reflection. Therefore, the geodesics with $r_c = r_g + 0$ and $E \to +0$ leave the black hole quadrant through the $\bar v < 0$, $\bar u > 0$ quadrant, too; in the limit $E = 0$ the exit point tends to the centre $(\bar v, \bar u) = (0, 0)$. 

In the same manner geodesic $m_2^\mathrm{prior}$ is reflected to $m$, while the light line 
$\ell_2^\mathrm{prior}$ is reflected to light line $\ell$.

The question is what is an origin of geodesics occupying the cone of past in the sector from $\alpha$ clockwise to $\ell_2^\mathrm{prior}$. Are such geodesics coming from singularity at $r=0$?

\begin{figure}[t]
\begin{center}
\includegraphics[width=9cm]{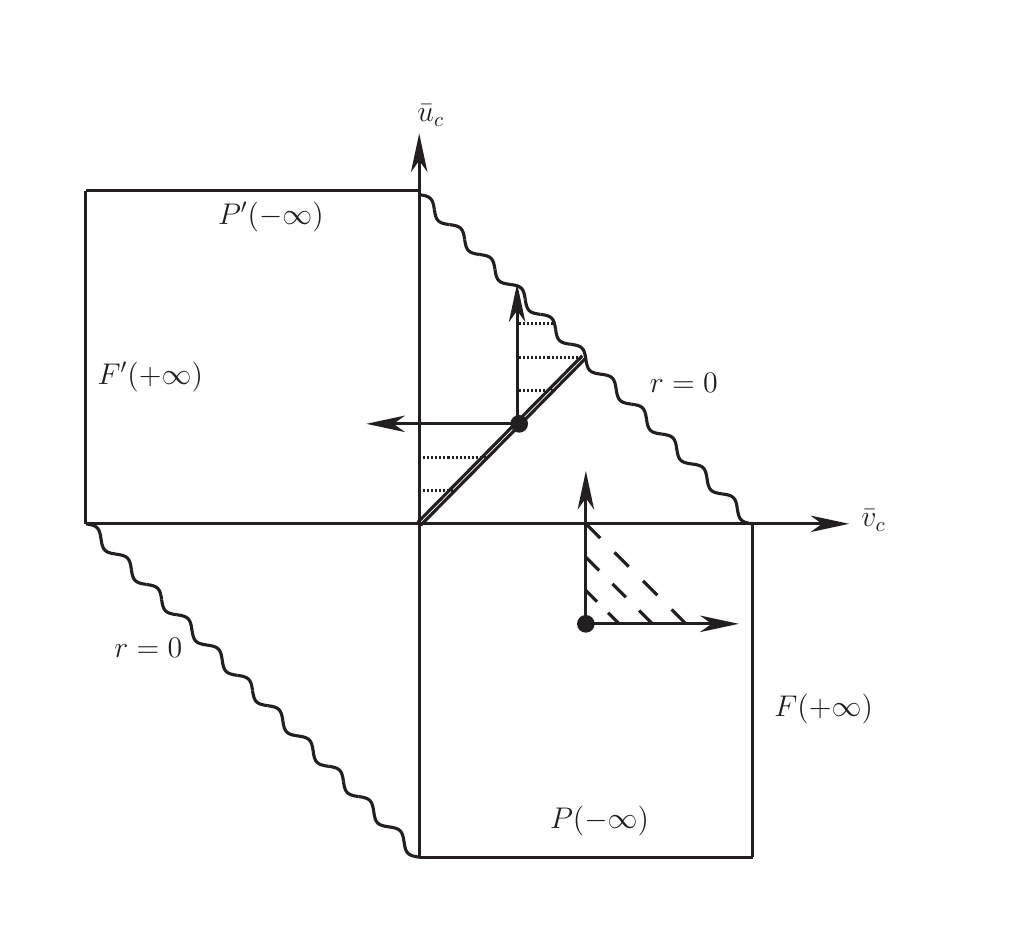}
\caption{The compact presentation of four domains of space-time in $(\bar v_c,\bar u_c)$-plane. The singularity $r=0$, two points with cones of future and infinity pasts and futures for distant observers are shown. The horizon coincides with the coordinate axes.}
\label{pic4}
\end{center}
\end{figure}

Consider a compactifying transform for Kruskal variables
\begin{equation}
\label{2in-9}
	\bar v_c=\tanh (\bar v/r_g),\qquad \bar u_c=\tanh (\bar u/r_g),
\end{equation}
so that differentials equal 
$$
	\mathrm d \bar v=r_g \cosh^2(\bar v/r_g)\cdot\mathrm d \bar v_c,\qquad
	\mathrm d \bar u=r_g \cosh^2(\bar u/r_g)\cdot\mathrm d \bar u_c.
$$
Therefore, according to (\ref{ds-uv}) the metric takes the conformal form
\begin{equation}
\label{ds-uv2}
	\mathrm d s^2=\frac{r_g^3}{r}\,\mathrm e^{-r/r_g}\,\cosh^{2}(\bar v/r_g)\cdot \cosh^{2}(\bar u/r_g)
	\cdot\mathrm d \bar v_c\,\mathrm d \bar u_c.
\end{equation}
Light lines with $\mathrm d s^2=0$ correspond to straight lines of $\mathrm d \bar v_c=0$ or 
$\mathrm d \bar u_c=0$. Then we can present the compact form of $(\bar v,\bar u)$-plane in terms of $(\bar v_c,\bar u_c)$ as Carter--Penrose diagram shown in FIG. \ref{pic4}. Horizon corresponds to two axes: $\bar v_c=0$ and $\bar u_c=0$. The infinity past for the distant observer is marked by $P(-\infty)$ line, and the infinity future is labeled by $F(+\infty)$ line. Analogous marks $P'(-\infty)$ and $F'(+\infty)$ are shown for the distant observer-prime, who lives in the opposite quadrant $\bar v < 0$, $\bar u > 0$. These marks agree with the Kruskal coordinates $\bar v$ and $\bar u$, in which the straight lines of constant time are ordered as noted after \eqref{eq:reflection}.

Two dots with cones of future are pictured over the horizon and under the horizon. We do not show the cones of past for a clarity of FIG. \ref{pic4} (see details in FIG. \ref{pic1} and FIG. \ref{pic2}), but the mentioned question about the past behind the horizon remain to speculate: a half in the cone of past behind the horizon comes from the singularity. 

\subsection{Internal horizon}
\begin{figure}[t]
\begin{center}
\includegraphics[width=10cm]{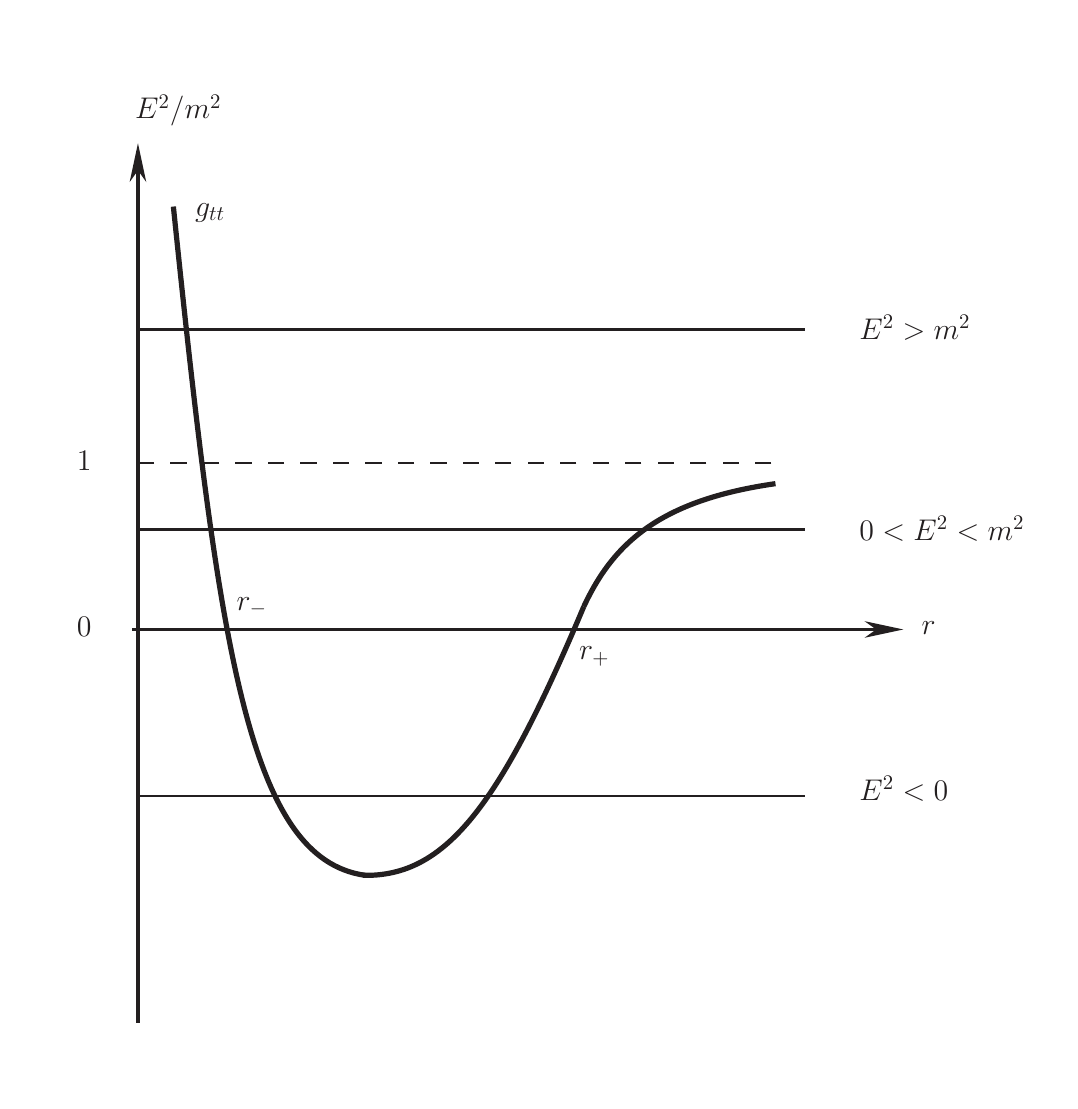}
\caption{Different levels of motion invariant in (\ref{rs2}).}
\label{pic6}
\end{center}
\end{figure}

Let us start with Carter--Penrose diagram for the Reissner--Nordstr\o m black hole, that possesses two horizons, the external at $r_+$ and internal at $r_-<r_+$ due to an electric charge $Q\neq 0$, while the limit of $Q\to 0$ results in $r_-\to 0$ standing for Schwarzschild black hole under study. So, 
\begin{equation}
\label{rs1}
	g_{tt}=-g_{rr}^{-1}=\frac{1}{r^2}\,(r-r_+)(r-r_-),\quad M=\frac12(r_++r_-),\quad Q^2=r_+ r_-.
\end{equation}
Here and below $G = 1$; the two horizons are distinct for $0 < Q^2 < M^2$, and in the limit $Q \to 0$ one has $r_+ \to r_g = 2M$ and $r_- \to 0$. The tortoise radius in accordance with $\mathrm d r_\star=\mathrm d r/g_{tt}$ takes the form
\begin{equation}
\label{rstar}
	r_\star=r+\frac{r_+^2}{r_+-r_-}\ln\left(\frac{r}{r_+}-1\right)- 
	\frac{r_-^2}{r_+-r_-}\ln\left(\frac{r}{r_-}-1\right).
\end{equation}
Radial geodesics possess the integral of motion $E$ and analogue of (\ref{in8}) 
\begin{equation}
\label{rs2}
	\frac{E^2}{m^2}=g_{tt}+\frac{1}{m^2}\big(\partial_{r_\star}\mathscr S\big)^2,
\end{equation}
that results in picture of levels shown in FIG. \ref{pic6} similar to FIG. \ref{H-M-fig}. However, massive particles never reach $r=0$: for $Q \neq 0$ one has $g_{tt} \to +\infty$ there, so the kinetic term $E^2/m^2 - g_{tt}$ cannot stay positive near $r = 0$. 
The derivative determining the velocity, the quantity
\begin{equation}
\label{rs2a}
	\left(\frac{\mathrm d t~}{\mathrm d r_\star}\right)^2=\frac{E^2/m^2}{E^2/m^2-g_{tt}}
\end{equation}
remains positive for $E^2 > 0$, while for $E^2 < 0$ it is negative and $\mathrm{d}t$ is imaginary. At the marginal trajectory with $E \to +0$ again $\mathrm{d}t/\mathrm{d}r_\star = 0$.

Relevant Kruskal coordinates are defined as 
\begin{equation}
\label{2in-1a}
	\left\{\begin{array}{l}
	u = t-r_\star,\\ v=t+r_\star,
	\end{array}
	\right.
\end{equation}
setting barred variables at $r_-<r<\infty$ 
\begin{equation}
\label{rs3}
	\left\{\begin{array}{l}\displaystyle 
	\bar u_+ = -2 \frac{r_+^2}{r_+-r_-}\,\exp\left(-\frac{u}{2r_+^2}(r_+-r_-)\right)\cdot C_+,\\ [5mm] \displaystyle 
	\bar v_+ = +2 \frac{r_+^2}{r_+-r_-}\,\exp\left(+\frac{v}{2r_+^2}(r_+-r_-)\right)\cdot C_+,
	\end{array}
	\right.
\end{equation}
and at $0<r<r_+$ 
\begin{equation}
\label{rs4}
	\left\{\begin{array}{l}\displaystyle 
	\bar u_- = +2 \frac{r_-^2}{r_+-r_-}\,\exp\left(+\frac{u}{2r_-^2}(r_+-r_-)\right)\cdot C_-,\\ [5mm] \displaystyle 
	\bar v_- = -2 \frac{r_-^2}{r_+-r_-}\,\exp\left(-\frac{v}{2r_-^2}(r_+-r_-)\right)\cdot C_-,
	\end{array}
	\right.
\end{equation}

where the constants $C_\pm$ absorb the phases of the logarithms in \eqref{rstar} when the two maps are matched on the overlap $r_- < r < r_+$ (see \cite{Birrell:1982ix,Soltani:2023tfr,Cvetic:2018dqf}). They cancel in the metric \eqref{rs5}.  The outer horizon $r_+$ is given by $\bar v_+=0$ and $\bar u_+=0$ axes, while $r_-$ is not covered by $(\bar v_+,\bar u_+)$-map. The inner horizon $r_-$ is given by $\bar v_-=0$ and $\bar u_-=0$ axes. The metric takes the conformal form
\begin{equation}
\label{rs5}
	\mathrm d s^2=-g_{tt}\frac{4r_{g+}^2}{\bar v_+\bar u_+}\,\mathrm d \bar v_+\mathrm d \bar u_+
	=-g_{tt}\frac{4r_{g-}^2}{\bar v_-\bar u_-}\,\mathrm d \bar v_-\mathrm d \bar u_-,
\end{equation}
where
$$
	r_{g+}=\frac{r_+^2}{r_+-r_-},\qquad r_{g-}=\frac{r_-^2}{r_+-r_-},
$$
Light geodesics are given by straight lines. The trajectories of massive particles satisfy $\mathrm d s^2>0$. The marginal geodesics at $E\to +0$ are presented by straight lines passing through the point $(\bar v_+,\bar u_+)=(0,0)$ the bifurcation point of the outer horizon, and ending on the curves $r = r_-$.
In the region $r < r_-$ massive particles have ordinary turn points, and light rays are reflected at $r = 0$ by the same rule, so that geodesics of both kinds go from the past $P$ to the future $F''$.

\begin{figure}[t]
\begin{center}
\includegraphics[width=11cm]{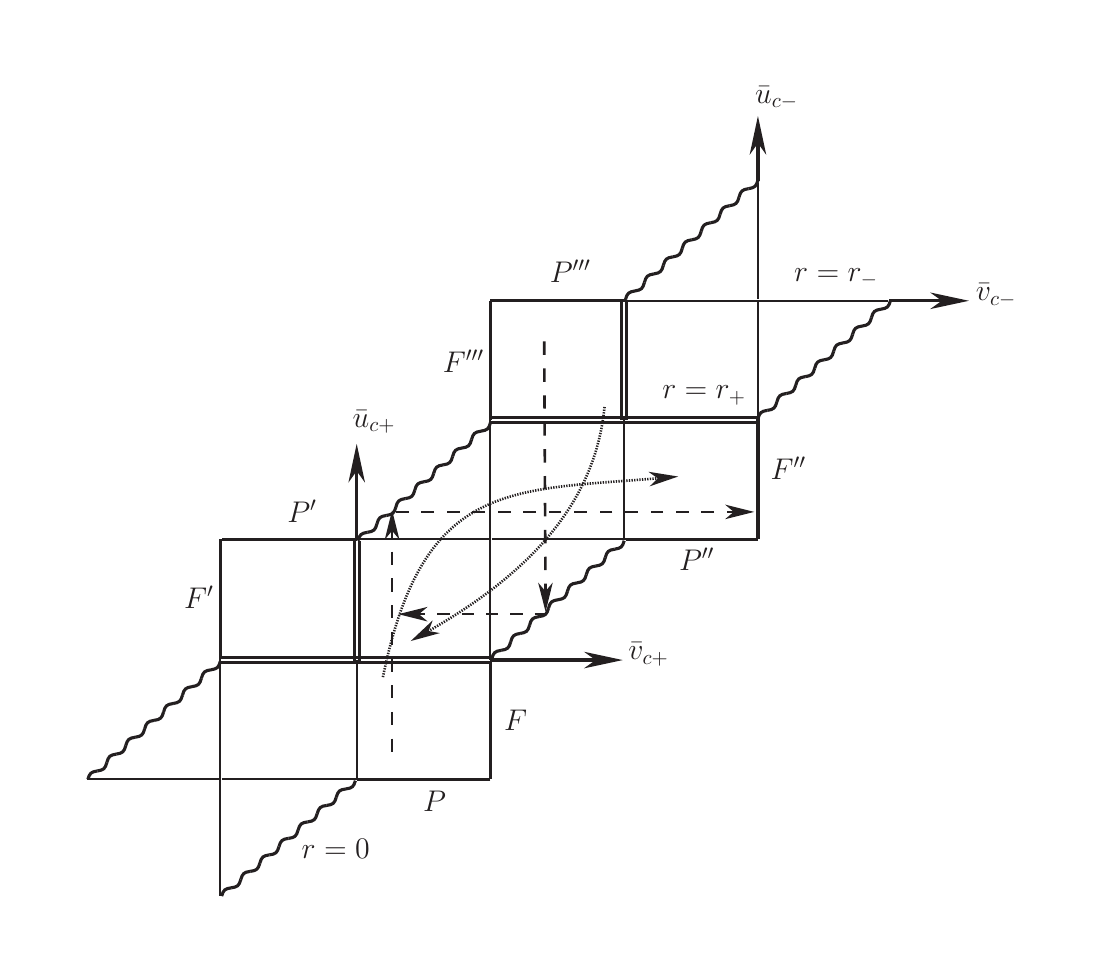}
\caption{The band of causal structure for the space-time of charged Reissner--Nordstr\o m black hole. The horizon $r=r_+$ is shown by a double line, while $r_-$ is pictured by thin line. Light geodesics are dashed, while trajectories of massive particles are marked by dotted curves.}
\label{pic7}
\end{center}
\end{figure}

The compact version of causal structure of space-time is presented in FIG. \ref{pic7}. In the region of $r<r_-$ we see the ordinary turn point, so that for instance geodesics of both light and massive particles go from the past $P$ to the future $F^{\prime\prime}$. Thus, no geodesics will come back into the $P-F$ quadrant. Geodesics do not form cycles. 

\begin{figure}[t]
\begin{center}
\includegraphics[width=10cm]{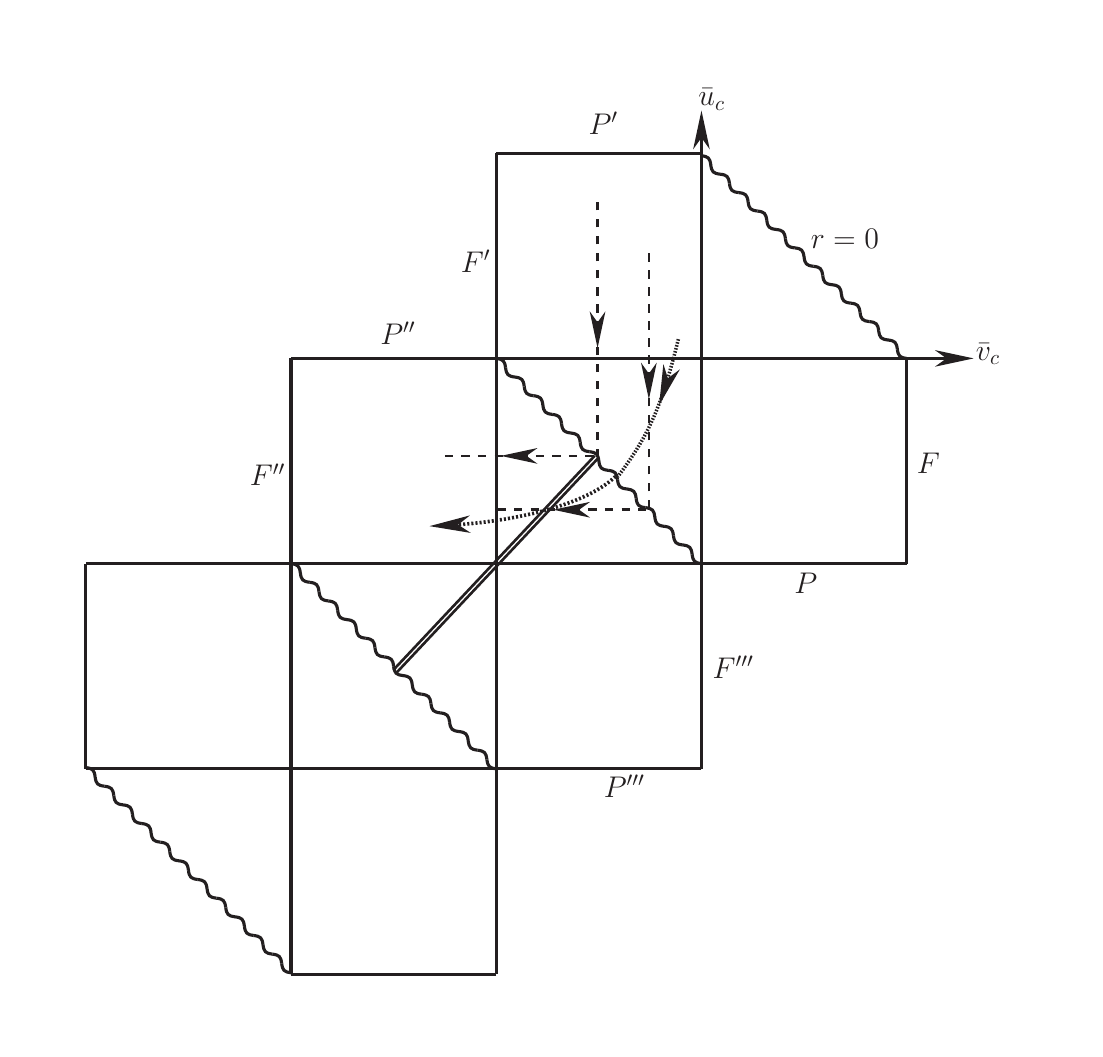}
\caption{The band of causal structure for the space-time of Schwarzschild black hole, the singularity $r=0$ is pictured by wave-curve. Boundaries of past and future for different distant observers are marked by $P$ and $F$ symbols with appropriate primes. Trajectories coming from the domain of distant observer-prime are reflected by the  singularity to the domain of distant observer-double prime, and they cannot be directed to the domain of initial distant observer. Geodesics of light are marked by dashed lines, while the geodesic of massive particle is shown by the dotted curve. The marginal geodesic of $E=0$ is presented by the double line. The singularity and horizons are multiple times  copied.}
\label{pic5}
\end{center}
\end{figure}

If we consider the limit of $r_-\to 0$, then the band of space-time structure does not disappear. Connected regions restricted by $r=0$ (wavy line) and $r=r_-$ (thin lines) in FIG. \ref{pic7} degenerate into wavy line of $r=0$. For any $Q \neq 0$ the singularity is timelike and becomes spacelike only in the limit, so the band in FIG. \ref{pic5} is understood as the limit of the diagram in FIG. \ref{pic7}. This results in the band of causal structure for the space-time of Schwarzschild black hole shown in FIG. \ref{pic5} (in this figure we change the prime marks for times of past and future). This picture clarifies the nature of geodesics in the cone of future propagating to the future $F^{\prime\prime}$. So, light lines from the past marked by $P^\prime$ reach the singularity after which the geodesics are reflected not to the future marked by $F$, but to $F^{\prime\prime}$. We show also the marginal trajectory with $E=+0$ in FIG. \ref{pic5}. Then a trajectory for a massive particle as marked by dotted curve occupies the correct path between the light and marginal curve symbolically shown as double straight line.

Therefore, the structure of the cones of future and past obtained by the reflection method is consistent with the band of the Reissner--Nordstr{\o}m space-time in the limit $r_- \to 0$. 

Other arguments for the infinite band of causal structure for space-time of the Schwarzschild black hole were given in \cite{Zeng:2025lch,Zeng:2023ueq} considering the radial motion in a quantum mechanical manner. So, rephrasing the Hamilton--Jacobi equation of (\ref{in1}) in terms of the 4-velocity
$$
	u^\mu=\frac{\mathrm d x^\mu}{\mathrm d \tau},\qquad u^\mu=\frac{1}{m}\,p^\mu
	=\frac{g^{\mu\lambda}}{m}\,\frac{\partial S}{\partial x^\lambda},
$$
one gets 
\begin{equation}
\label{in1-a}
	g_{\mu\nu}\,u^\mu u^\nu=1.
\end{equation}
Substituting 
$$
	\frac{\partial S}{\partial t}=-E,\qquad g^{tt}=g_{tt}^{-1},\qquad g_{rr}= - g_{tt}^{-1}
$$
results in the following:
$$
	g_{tt}\,\left(-g_{tt}^{-1}\,\frac{E}{m}\right)^2-g_{tt}^{-1}\,
	\left(\frac{\mathrm d r}{\mathrm d \tau}\right)^2=1\quad \Rightarrow\quad
	\left(\frac{\mathrm d r}{\mathrm d \tau}\right)^2+g_{tt}=\frac{E^2}{m^2}.
$$
For the Schwarzschild metric we use $m\,\mathrm d r/\mathrm d\tau\equiv \hat p_r$ and arrive at 
\begin{equation}
\label{eq-chine}
	\frac{1}{2m}\,\hat p_r^2+\frac{m}{2}\left(1-\frac{2GM}{r}\right)=\frac{m}{2}\,\frac{E^2}{m^2}
	\quad \Rightarrow\quad
	\frac{1}{2m}\,\hat p_r^2- \frac{GMm}{r}=\frac{m}{2}\left(\frac{E^2}{m^2}-1\right),
\end{equation}
which is the equation of radial motion in the Coulomb-like field of attraction. In \cite{Zeng:2025lch,Zeng:2023ueq} the quantity $\hat p_r$ is treated as the quantum mechanical operator, acting on a stationary wave-function for a particle of mass $m$. Then one can treat bound levels at $E<m$. In such a treatment the author of \cite{Zeng:2025lch,Zeng:2023ueq} concludes that, because of the flow of proper time, the causal structure of the black hole is not restricted to a single segment of the band.

So, the infinite band structure of causal space-time regions for the radial motion with energy $E>0$ is certainly established. Our next item to study concerns time-like geodesics with the conserved parameter $r_c<r_g$.

\section{Maps for Interior Geodesics\label{sec-nonH}}
Time-like trajectories of massive particles lying completely in the interior of the Schwarzschild black hole have got conserved value $r_c$ in the region $0<r_c\leqslant r_g$ and possess $({\mathrm d t}/{\mathrm d r_\star})^2\leqslant 0$ in accordance with (\ref{in13}). So, the only opportunity to define  real-valued Kruskal variables corresponds to $r_c=r_g$, when $({\mathrm d t}/{\mathrm d r_\star})^2\equiv 0$ and (\ref{2in-6}) is valid: a straight line with constant slope equal to $\varkappa^2>0$ passes from $r=0$ to the center of $(\bar v,\bar u)$-plane (see FIG. \ref{pic2} and FIG. \ref{pic4}) and
\begin{equation} 
\label{2in-6a}
	\frac{\mathrm d \bar u}{\mathrm d \bar v}=\frac{\bar u}{\bar v}\equiv \varkappa^2.
\end{equation}
This fact of marginal geodesics with $E = 0$ indicates a possibility to construct a non-Hausdorff extension of space-time, where these geodesics serve as the one-dimensional  curve of touching between two two-dimensional manifolds: the real $(\bar v, \bar u)$-plane and the disc $(\varrho, \varphi_\textsc{e})$ introduced below. Let us set 
\begin{equation}
\label{map1}
	\left\{\begin{array}{l}\displaystyle 
	\bar u = -\varkappa\,\varrho\, \mathrm i\cdot\exp\left(\mathrm i\varphi_\textsc{e}\right),\\ [3mm] 
	\displaystyle 
	\bar v = +\frac{1}{\varkappa}\,\varrho\, \mathrm i\cdot\exp\left(-\mathrm i\varphi_\textsc{e}\right),
	\end{array}
	\right.\qquad \Rightarrow\quad \bar u\cdot \bar v=\varrho^2,\quad 
	\bar u =-\varkappa^2\,\bar v\,\exp\left(\mathrm i2\varphi_\textsc{e}\right),
\end{equation}
where $\varphi_\textsc{e}\in [0,2\pi]$ and 
for the marginal geodesics one has to put $\varphi_\textsc{e}=\pi/2$ or 
$\varphi_\textsc{e}=3\pi/2$: for these two choices the geodesics belong to $(\bar v>0,\bar u>0)$-quadrant and $(\bar v<0,\bar u<0)$-quadrant, correspondingly, if we fix the sign of parameter $\varkappa>0$ and $\varrho\geqslant 0$. Since the interior is restricted by the singularity and 
$\bar u\cdot \bar v\leqslant 4r_g^2$, we deduce
$$
	\varrho\in [0,2r_g].
$$
Here $\varrho=0$ stands for the horizon, while $\varrho=2r_g$ does for the singularity. 

\begin{figure}[t]
\begin{center}
\includegraphics[width=8cm]{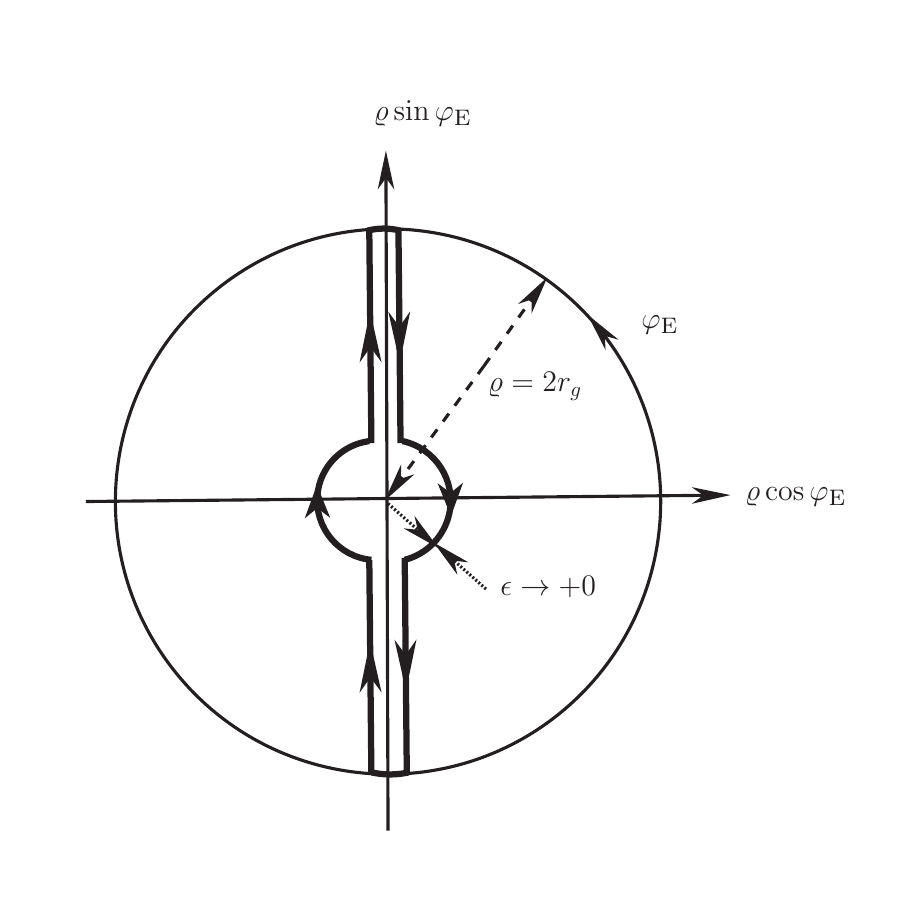}
\caption{Disc of black hole interior with the marginal geodesic of energy $E=0$ ($r_c\to r_g$). The path runs along $\varphi_\textsc{e} = \pi/2$ in the black hole and along $\varphi_\textsc{e} = 3\pi/2$ in the white hole, with the reflection at $r = 0$ and semicircles of radius $\epsilon \to +0$ around the horizon point.}
\label{pic8}
\end{center}
\end{figure}

The marginal geodesics with $r_c=r_g$ can be pictured in the polar coordinates of two-dimensional disc as shown in FIG. \ref{pic8}. We use (\ref{eq:kruskal}) to express time and tortoise radius after (\ref{map1})
\begin{equation}
\label{eq:kruskalb}
	\left\{\begin{array}{ll}
	t &\hskip-2.mm=\displaystyle  r_g\,\ln\left(-\frac{\bar v}{\bar u}\right),\\[5mm]  
	r_\star &\hskip-2.mm=\displaystyle  r_g\,\ln\left(-\frac{\bar v\cdot \bar u}{4r_g^2}\right),
	\end{array}
	\right.
	\qquad \Rightarrow\qquad
	\left\{\begin{array}{lll}
	t &\hskip-2.mm=\displaystyle  -\mathrm i\,2\,r_g\,\varphi_\textsc{e}+\Delta t_0,
	& \qquad\Delta t_0=-2 r_g\ln\varkappa,\\[5mm]  
	r_\star &\hskip-2.mm=\displaystyle  2\,r_g\,\ln\left(\frac{\varrho}{2r_g}\right).
	& 	\end{array}
	\right.
\end{equation} 
So, the full period of evolution in terms of $\varphi_\textsc{e}$ transforms to the period in imaginary time $\beta= 4\pi r_g$ setting the inverse temperature if one considers this motion as stationary thermal ensemble. 

The direction of loop in FIG. \ref{pic8} is not important at $E=0$. The pattern of motion in FIG. \ref{pic8} is significant, since if one restricts the consideration by  the $(\bar v, \bar u)$-plane separately, then one can find the half value of period in imaginary time $\beta/2$ by putting the turn point at the horizon and reflection point at $r=0$ which makes the cycle in the single quadrant instead of two which are necessary in the disc coordinates. This may be the reason for doubling of the black hole temperature $T_\textsc{bh}=1/\beta$ in \cite{tHooft:2024auh} by G.\,'t Hooft.

From expression (\ref{ds-uv}) for metric in terms of $\bar v$ and $\bar u$ and following the transform in (\ref{map1}) we find the metric
\begin{equation}
\label{ds-uv-a}
	\mathrm d s^2=\frac{r_g}{r}\,\mathrm e^{-r/r_g}
	\cdot\mathrm d \bar v\,\mathrm d \bar u= 
	\frac{r_g}{r}\,\mathrm e^{-r/r_g}
	\cdot(\mathrm d \varrho^2+\varrho^2\mathrm d \varphi_\textsc{e}^2).
\end{equation}
The metric is Euclidean, as expected for the imaginary time. The conformal factor is positive, and time-like geodesics inside the interior of the black hole are meaningful at $0<r_c< r_g$, too. One can suggest transitions from one integral of motion $r_c\geqslant r_g$ to another $r_c\leqslant r_g$. The question is what is a concept of states with $r_c<r_g$ should be.

\subsection{Thermal Island}
While positive values $r_c<r_g$ are real-valued, the <<energy>>  $E$ and <<time step>> $\mathrm d t$ are imaginary, certainly. The latter indicates the thermal evolution. But what is the energy of the system at $\mathfrak{Im} E\neq 0$? 

One has to assume that for each state with $E_+=\mathrm i \mathscr E$ at real value $\mathscr E\in \mathbb{R}$ another state with $E_-=-\mathrm i \mathscr E$ should be introduced in some coherent sense. In that case the total balance of energy remains real and equal to zero, as if there is nothing inside the disc $(\varrho,\varphi_\textsc{e})$ of the island in the black hole interior  at all. 

Some pure quantum states are excluded since any interference breaks the thermal description. So, one has to introduce a static density matrix $\hat\rho$ for two states $|E_+\rangle$ and $|E_-\rangle$, which  takes the form
\begin{equation}
\label{island1}
	\hat \rho =w_\mathscr{E}\left(\begin{array}{cc} 
	|E_+\rangle\,\langle E_+| & 0 \\[1mm]
	0 & |E_-\rangle\,\langle E_-| \end{array}\right),
\end{equation}
where probability $w_\mathscr{E}$ is setting Gibbs-like distribution. 
 Some speculations in this avenue can be found in \cite{PhysRevD.72.124011,Kiselev:2004vx,Kiselev:2005yz,Kiselev:2005ar,Kiselev:2013ora}. 
 In this way the spectrum of $\mathscr E$ is discrete because of periodic conditions in $\varphi_\textsc{e}$. At this moment it is significant that the thermal island arises in this conditions, and it is common for all domains in the band of causal structure for the space-time of Schwarzschild black hole. 

Zero balance of energy in the island with account for the opposite direction of motion versus the evolution parameter $\varphi_\textsc{e}$ and mixed form of density matrix for two states with opposite imaginary energies signalise that the stress-energy tensor vanishes identically at regular points of the $(\bar v,\bar u)$-domains. Therefore, particles in the island do not break the condition of empty matter in regular points for the black hole solution in general relativity. Moreover, the entropy of black hole can be identified with the thermal entropy of particles in the island, while the metrics field can be treated as the mean self-consistent field of these particles. If the motion is considered as a stationary thermal ensemble, the period $\beta = 4\pi r_g$ gives the temperature $T = 1/(8\pi G M)$. The first law $\mathrm{d}M = T\,\mathrm{d}S$ then yields the Bekenstein--Hawking entropy $S = 4\pi G M^2 = A/(4G)$, where $A = 4\pi r_g^2$ is the horizon area. A direct calculation of the island entropy from the density matrix \eqref{island1} remains to be done.

How can one describe the transition of particle with $E\geqslant 0$ to the island with positive $r_c\leqslant r_g$ in the interior of black hole? A plausible way involves the notion of entanglement between a particle state and a state in quantum gravity. This entanglement is a possible way to reach the mixed thermal state in the particle sub-sector, since the entanglement can originate the decoherence formed in the sub-sector. However, a description of such entanglement is not available at the moment, because no ultimate formulation of quantum gravity exists yet.

This island probably can be related to islands discussed in \cite{Morozov:2023jsp}. 

\section{Conclusion}
The notion of reflection for the geodesic motion allows us to confirm the causal structure of space-time for the domain of distant observer far from the black hole horizon and to precise the form of the cones of future and past behind the horizon. The study obtains the infinite band of causal domains for the Schwarzschild black hole as the limit of the Reissner--Nordstr{\o}m black hole, when the internal horizon tends to the singularity. 

The investigation shows the existence of geodesics confined in the black hole interior which can form the thermal bath due to transitions involving the entanglement between the particle and quantum gravity states under further decoherence in the particle sub-sector of states, which is expressed in terms of the density matrix for mixed particle states. Under the thermal interpretation, the period of imaginary time $\beta = 4\pi r_g$ gives the Hawking temperature, and the first law then leads to the area law for the entropy.

Consequences of introducing  of thermal island in the black hole interior require further comprehensive investigations. 


\bibliography{bib_BH-interior}
	
\end{document}